%% file: Dirac_vaxjo2010.tex
\documentclass[final,numberedheadings]{aipproc}
\usepackage{amssymb}
\usepackage{ccaption}
\input{myqcircuit} \layoutstyle{6x9} \indentcaption{1.2em}
\captionnamefont{\bf}\captiondelim{\bf.~}
\captiontitlefont{\rmfamily}\precaption{\hskip 1em\footnotesize {\bf FIGURE~}}
\def\eg{{e.~g.} }\def\ie{{i.~e.} }\def\paragraph#1{\medskip\par\noindent{\bf #1}}
\def\lntrnsfrm#1{\mathrm #1} 

\def\rA{\lntrnsfrm A}\def\rB{\lntrnsfrm B}\def\rC{\lntrnsfrm C}

\def\<{\langle}\def\>{\rangle}

\renewcommand{\geq}{\geqslant}\renewcommand{\leq}{\leqslant}

\def\<{\langle}\def\>{\rangle}

\def\Hg#1{{\rm H^{(#1)}_{gate}}}

\begin{document}
\title[]{Physics as Information Processing \footnote{Work presented at the conference {\em Advances
      in Quantum Theory} held on 14-17 June 2010 at the Linnaeus University, V\"axj\"o, Sweden.}}
 \author{Giacomo Mauro D'Ariano}{address={{\em QUIT} Group,
    Dipartimento di Fisica ``A. Volta'', 27100 Pavia,
    Italy, {\em http://www.qubit.it}\\
    Istituto Nazionale di Fisica Nucleare, Gruppo IV, Sezione di Pavia}}
\begin{abstract} 
  I review some recent advances in foundational research at Pavia QUIT group.  The general idea is
  that there is only Quantum Theory without quantization rules, and the whole Physics---including
  space-time and relativity--is emergent from the quantum-information processing. And since Quantum
  Theory itself is axiomatized solely on informational principles, the whole Physics must be
  reformulated in information-theoretical terms: this is the {\em It from Bit} of J. A. Wheeler.

  The review is divided into four parts: a) the informational axiomatization of Quantum Theory; b)
  how space-time and relativistic covariance emerge from quantum computation; c) what is the
  information-theoretical meaning of inertial mass and of $\hbar$, and how the quantum field
  emerges; d) an observational consequence of the new quantum field theory: a mass-dependent
  refraction index of vacuum. I will conclude with the research lines that will follow in the
  immediate future.
\end{abstract}
\maketitle
%%%%%%%%%%%%%%%%%%%%%%%%%%%%%%%%%%%%%%%%%%%%
%% MAINMATTER
%%%%%%%%%%%%%%%%%%%%%%%%%%%%%%%%%%%%%%%%%%%%
\section{Introduction}
After more than a century Quantum Theory (QT) has never been falsified in any experiment, and
maintains its unprecedented predicting power. However, the derivation of its mathematical framework
from practical principles has remained an open problem. Stimulated by the work of Lucien Hardy---who
gave a first set of simple principles for QT \cite{Hardy}---and by Chris Fuchs---who argued that QT
is a special kind of information {\em ``and only a little more''} \cite{Fuchs}---in the summer of
2003 I started my axiomatization program. My dissatisfaction with Lucien's principles was that they
were still of mathematical nature, whereas, as regards Chris, I couldn't imagine what was his {\em
  little more}. I presented the first very preliminary results of the program at the 2005 Vaxjo
conference \cite{my2005}, introducing the operational axiomatization framework and providing a first
list of candidate principles. In the recent article \cite{myCUP2009} I reached a stable set of
purely operational postulates, though still incomplete. QT was there regarded as a set of rules that
are sufficient for the experimenter to make predictions on future events on the basis of suitable
tests. The postulates were: Causality, Local Observability, and the existence of Faithful States
that allow calibrations of instruments and local preparation of joint states.  Such postulates have
shown an unprecedented power, excluding all known probabilistic theories except QT.  The
axiomatization program later continued in collaboration with my former students G.  Chiribella and
P.  Perinotti. In Ref.  \cite{CDP2010} we have discovered the full potential of a new Purifiability
postulate, and started afresh with a new approach that provides a diagrammatic way of proving
theorems. From the postulates we derived essentially all the relevant features of QT and Quantum
Information, including dilation and error correction theorems, teleportation, no-cloning, and more.
In the fall of 2009 we reached a complete set of postulates, that have been under scrutiny for the
entire last year, and which seem to have finally reached the state of a manuscript
\cite{ournature,support} (of the two references the former is intended to be a popularization,
whereas the latter contains the precise formulation and the mathematical proofs).  QT is derived
from six principles that can be entirely formulated in the common language of computer programming:
the original probabilistic operational approach is indeed nothing but the most general
information-processing framework.  We still don't know if the set of postulates is minimal (logical
independence of axioms is knowingly an hard problem): indeed, as mentioned, just Causality, Local
Discriminability and Purifiability are sufficient to derive all the relevant informational features
of QT. Nevertheless, the new postulates are essential in the mathematical derivation. I will briefly
review the postulates in Sect.  \ref{Qinf}. Let me mention here that among the Lucien's {\em five
  reasonable axioms} \cite{Hardy} Causality was completely overlooked: the postulate is indeed so
natural that remained hidden in the framework. However, as shown in Refs.  \cite{myCUP2009,CDP2010},
it is easy to build up a theory similar to QT without causality, and the relevance of recognizing
Causality as a postulate is obvious in view of possible non-causal variations of QT for a theory of
Quantum Gravity.

At the beginning of October 2009 I realized that the informational axiomatization program could go
well beyond the narrow boundaries of QT and involve Relativity and Quantum Field Theory (QFT), \ie
the whole Physics. The point is: What is QFT more than just an infinite set of quantum systems
locally interacting, \ie a huge quantum computer? But then, Relativity and space-time itself should
be emergent features of the quantum computation. In other words, Special Relativity must be derived
from Causality.

The idea of deriving the geometry of space-time from a purely causal structure (such as a quantum
computer) has been a long-term program initiated by Rafael Sorkin and collaborators
\cite{Bombelli-Sorkin_(1987)} more than two decades ago! However, I wanted to see how the Lorentz
transformations can be derived from causality. Remembering the Tomonaga-Schwinger foliation in their
approach to QFT, I recognized the strict similarity with the foliation that Blute, Ivanov, and
Panangaden \cite{Blute} built up on causal networks (Lucien was perceiving the same point at the
same time \cite{Hardy:2009}). I whence realized that one should be able to {\em see} the Lorentz
transformation on the quantum circuit by building up two different global foliations and comparing
their synchronous sets of events. In this way I discovered that, upon introducing the metric as
simple event-counting, from the causal network space and time emerge already endowed with the
relativistic covariance. It was a startling experience to actually see the Lorentz time-dilation and
length-contraction on the quantum circuit!

In writing the conference proceedings of my 2009 talk in Vaxjo \cite{vax2009} I couldn't resist
reporting this idea (it will be briefly revisited here in Sect.  \ref{s:sremerg}). Five months after
I posted the manuscript on the web, I was happily surprised by the appearance on the web of the work
by Kevin Knuth and his student Newshaw Bahreyni \cite{Knuth}, where they showed how the Minkowski
signature can be derived from a causal poset. They indeed also claimed the derivation of the Lorentz
transformations, but after careful reading their manuscript---indeed very interesting in
consideration of the generality of their notion of ``event''---I couldn't understand their
derivation.  In addition, I was unable to find the origin of what looked as a trivial error---the
velocity at power four instead of two in the Lorentz transformation. In the meanwhile, with my
student A.  Tosini we worked out an analytical derivation of the Lorentz transformations
\cite{Lorentz} based on the idea of Ref.  \cite{vax2009}, using a topological homogeneous causal
network. Later I asked Kevin about the apparent mistake in his manuscript, and it turned out that
its fixing needed a change of the whole derivation---the good news was that, as a byproduct, the new
derivation seemed to require the notion of spin! I still have not analyzed the second version of
Kevin's manuscript (I'm still waiting for a self-contained forthcoming version), but in my opinion
there is a crucial ingredient in the derivation of the Lorentz transformations---which still I don't
understand how Kevin can avoid---\ie the assumption of uniform topology of the causal network,
corresponding to the Galileo principle: we will come back to this point later. But, let me say:
everything looked so damned interesting!  Later I discovered other related very interesting works by
other authors in much earlier literature, such as Stephen Wolfram's idea that the Lorentz
transformations might be derived from a cellular automaton evolution \cite{WolframBook}, Leslie
Lamport's work \cite{lamport} on the problem of time-ordering of events in distributed systems, and,
{\em dulcis in fundo}, a very interesting derivation of the Lorentz composition of velocities based
on a simple random-walk in the book of Irving Stein\cite{stein}---who, at the antipodes of the
informational approach, seeking an {\em ontology} for Physics concludes that it is non classical!

Thus, after years of research spent in the tunnel of pure quantum theory, I finally saw the outdoor
light of ``space'' and ``time'', and experienced the joy of entering the territory of Relativity and
QFT. Under enthusiasm, I immediately tried to see what could be said about QFT: the program of
translating the whole Physics into information-theoretic terms looked huge and very ambitious. The
first question that I was able to address was: what is the informational meaning of mass? The answer
was simple: the Dirac's Zitterbewegung. Indeed, the inertial mass is nothing but the slowing-down of
the information flow due to repeated changes of direction. In this way one can define mass in terms
of the coupling between left and right-propagating information flows, which is expressed in terms of
the Compton wavelength. The statement of the equivalence of this notion of mass with the usual
notion (in the conventional units) leads also to recover $\hbar$ simply as the conversion factor
between the two! I will briefly report about this point in Sect. \ref{s:mass}. I liked this very
much, since I always wanted to make Quantum Mechanics independent from $\hbar$, the notion of
particle, and the quantization rule. Why? For many reasons. First: I never liked the quantization
rules: why we should start from the theory of Classical Mechanics---that we know is incorrect---to
derive QT---which we know is correct? Second: the notion of particle as an ontology becomes
inconsistent with the wave-particle dualism. Third: the notion of particle becomes totally
evanescent as an ontology in QFT, where it is just a quantum state---a state of the field! I wanted
to have the three things---$\hbar$, notion of particle, and quantization rules---as emergent.

Again under the enthusiasm for the last findings, I immediately tried to work-out a set of toy
models for the new Computational QFT, and rushed to write them on the same Vaxjo
proceedings\cite{vax2009}. The result was that I made my own mistake: in order to make calculations
easy I considered a quantum computer with gates infinitesimally close to the identity... without
realizing that, in this fashion I couldn't derive Relativity anymore from the
information-processing, since the maximal information speed must have been infinite!  (I hope that
the reader will forgive me: proceedings sometimes can be mistaken when they present work in
progress). I later understood that, on the contrary, the new Quantum Computational Field Theory
(QCFT) was very different from the usual quantum-computer simulation of Quantum Mechanics (see \eg
\cite{Bialynicki,Boghosian}), since it must have all unitary transformations very far from the
identity, \eg unitary {\em swaps} describing an information flow in a fixed direction at the maximum
causal speed of 1 gate/1 step. QCFT must be something closer to the quantum cellular automaton of
Ref.  \cite{Meyer}. And in order to recover the continuum we should not take the limit of vanishing
$a$ and $\tau$, but take instead the thermodynamic limit of infinitely many gates.

One of the good news about the new QCFT, is that it has observational consequences. For example, the
request of unitariety for an evolution linear in the field leads to a mass-dependent vacuum
refraction index. As I will show in Sect. \ref{refrindx}, this phenomenon is simply a consequence of
the normalization of the rows of the unitary matrix giving the linear evolution of the field: if we
increase the mass (\ie the coupling between modes), then we need to decrease the maximal propagation
speed. This is a general feature of QCFT, and is due to the fact that QCFT respects causality,
namely no information can propagate at a speed greater than $c$.

\section{Quantum Theory as Information Theory\label{Qinf}}
In this section I will briefly review the six informational principles for QT of Ref.
\cite{ournature}. As mentioned, the principles can be formulated in the common language of computer
programming. We just need to add few additional specifications. We call {\bf information processing}
the equivalence class of {\bf subroutines} that achieve the same input-output relation. The
subroutine corresponds to what it is usually called {\bf event} in the language of operational
probabilistic theories of Refs. \cite{myCUP2009,CDP2010}, whereas the equivalence class of the
information processing corresponds to the notion of {\bf transformation}. These are represented 
in form of a box with wires as follows
$$\Qcircuit @C=.8em @R=.8em { \ustick{\rA} & \multigate{1}{\mathcal{T}} &\ustick{\rA'}\qw\\
  \ustick{\rB} & \ghost{\mathcal{C}} & \ustick{\rB'}\qw}$$ where the left/right wires represent
input/output registers on which information is read/written, and different letters denote different
types of register. The {\bf register} corresponds to the notion of {\bf system} in the operational
framework.  We can compose processings connecting input with outputs of the same type as follows
$$\Qcircuit @C=.8em @R=.8em {
  \ustick{\rA}&\qw & \multigate{1}{\mathcal{T}_1} &\qw &\qw &\ustick{\rA'} \qw \\
  & & \pureghost{\mathcal{T}_1}    & \ustick{\rC} \qw&\multigate{1}{\mathcal{T}_2} & \\
  \ustick{\rB}& \qw & \qw &\qw &\ghost{\mathcal{T}_2} &\ustick{\rB'} \qw}$$ We should keep in mind
that the circuit representation is not a flow-diagram, but represents the entire run (\ie if we send
the output to the input of a previously called processing we will not draw a loop, but instead we
will redraw the same box twice). Therefore, precisely {\bf a box represents a single call of the
  processing}. The entire processing is thus described by a DAG (directed acyclic graph). For
subroutines and processings---as for events and transformations---it is possible to perform both
{\bf coarse-graining} and {\bf refinement}, \eg a subroutine can be divided into alternative
subroutines whose occurrence generally depends on the input (for example, the ``factorial'' we
can be divided into two alternatives---{\tt n=0} and {\tt n>0}---using the subroutine {\tt ``return
  1''} for $n=0$ and the subroutine {\tt ``return n*f(n-1)''} for {\tt n>0}).  The complete set of
alternative processings corresponds to the operational notion of {\bf test} (collection of events),
and is represented by a single box as follows
$$\Qcircuit @C=.8em @R=.8em { \ustick{\rA} & \multigate{1}{\{\mathcal{S}_i\}} &\ustick{\rA'}\qw\\
  \ustick{\rB} & \ghost{\{\mathcal{S}_i\}} & \ustick{\rB'}\qw}$$ We name the set of all possible
constituents of a processing its {\bf refinement set}, and call {\bf indivisible} a processing with trivial refinement
set. The data-input and data-output are themselves information processings: the
{\bf initialization} and {\bf readout}, respectively, corresponding to the operational notions of
{\bf state} and {\bf effect}. These are represented as follows
$$\begin{aligned}
  \Qcircuit @C=.8em @R=.8em {
    & \multiprepareC{2}{\{\mathcal{A}_i\}} &\ustick{\rA} \qw \\
    & \pureghost{\{\mathcal{A}_i\}}  & \ustick{\rB} \qw\\
    & \pureghost{\{\mathcal{A}_i\}} & \ustick{\rC} \qw}
\end{aligned}\qquad\qquad
\begin{aligned}\Qcircuit @C=.8em @R=.8em {
    &\ustick{\rA} \qw& \multimeasureD{2}{\{\mathcal{B}_i\}} \\
    & \ustick{\rB} \qw &\ghost{\{\mathcal{B}_i\}} \\
    & \ustick{\rC} \qw &\ghost{\{\mathcal{B}_i\}}}
\end{aligned}$$
Thus an ``indivisible initialization'' is just the same as a pure state. An initialization followed
by a processing can itself be regarded as a different initialization, namely
$$\begin{aligned}
  \Qcircuit @C=.8em @R=.8em { & \prepareC{\mathcal{A}} &\ustick{\rA}
    \qw & \gate{\mathcal{S}} & \ustick{\rB} \qw}
\end{aligned}\quad=\quad
\begin{aligned}
  \Qcircuit @C=.8em @R=.8em { & \prepareC{\mathcal{A}'} &\ustick{\rB}
    \qw}
\end{aligned}.$$ Finally, the {\bf domain} of a processing is the set of its possible
initializations, its {\bf range} is the set of its possible readouts. An initialization is {\bf
  specific} when its refinement set is not the whole set of initializations (in the operational
language this means that the state is on the boundary of the convex set). Two initializations
$\mathcal{A}_1$ and $\mathcal{A}_2$ are {\bf discriminable} when there exists a readout that occurs
with different probabilities on the two initializations, whereas the discrimination is {\bf perfect}
when the two probabilities are zero and one.

\smallskip
We are now ready to understand the postulates:

\smallskip
\begin{enumerate}
\item[\bf P1] {\bf Causality:} The occurrence probability of a component processing is independent
  on the choice of the processing at its output (i.~e. information flows from input to output, and
  not viceversa).
\item[\bf P2] {\bf Local Readability:} We can discriminate two
  initializations of multiple registers by readouts on single
  registers.
\item[\bf P3] {\bf Reversibility and Indivisibility of Computation:} Every information processing
  can be achieved through a reversible one, by adding auxiliary input registers in such a way that
  the joint initialization is indivisible.
\item[\bf P4] {\bf Indivisibility of Processing Composition:} The
  processing corresponding to the input-output sequence of two
  indivisible processings is itself indivisible.
\item[\bf P5] {\bf Discriminability of Specific Initializations:} For
  any specific initialization there exists another initialization that
  can be perfectly discriminated from it.
\item[\bf P6] {\bf Lossless Compressibility:} For any initialization
  there exists an encoding which is perfectly decodeable on its
  refinement set, and the encoded initialization is not specific.
\end{enumerate}

\medskip The six postulates involve different aspects of the information processing, and each one
generates a range of logical consequences that would require a long discussion: the reader is
addressed to Refs. \cite{ournature,support} for details. Here I just mention that Postulates P1, P2
and P4 correspond to Causality, Local-Discriminability, and Atomicity of Evolution of Ref.
\cite{myCUP2009}, respectively. P3 is the purification postulate of Ref.  \cite{CDP2010}. Postulates
P5 and P6 are new. I already briefly commented on P1 in the introduction. Regarding the other
postulates I just want to remind that P2 (Local Discriminability) is the origin of the complex
tensor product and reconciles holism with reductionism. P3 synthesizes the {\em parallelism} and
{\em reversibility} of quantum computation, and is the most ``quantum'' postulate, in the sense that
all postulates apart from P3 are satisfied by classical theory (P3 is not satisfied by PR boxes
\cite{toytheories}). P4 may look obvious: however, there is no reason why the same processing
obtained by composing two indivisible ones could not be itself achieved in principle by a subroutine
which is divisible. P5 and P6 also seem obvious, but it is easy to construct a theory that violates
P5, whereas P6 becomes non trivial in a general information-processing framework with different
types of registers. P6 is the starting point of Shannon's and Schumacher's compression.
I want to emphasize that currently no known theory satisfies P1, P2, and P3 apart from QT, and, as
shown in \cite{CDP2010}, P3 is the basis of most quantum information protocols. However, in the
derivation of QT in \cite{support} P5 and P6 play an essential role. The reader is addressed to
Refs.  \cite{ournature,support} for a thorough illustration of postulates and for the mathematical
derivation of QT from them.

\section{Emergence of Space-Time and Special Relativity from information processing\label{s:sremerg}}
\begin{figure}[ht]
\includegraphics[width=3.3in]{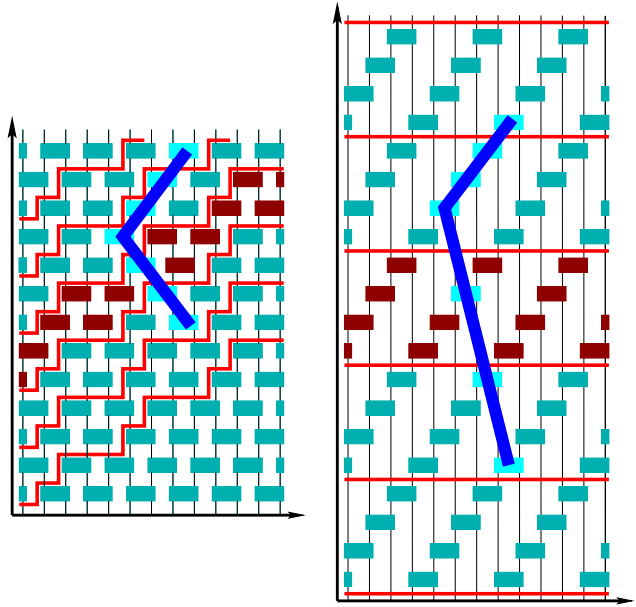}
\caption{\rm Illustration on how Special Relativity emerges from Local Causality. A chain of events
  representing a clock tic-tac (light bouncing between two mirrors) is highlighted in a
  topologically uniform computational circuit (on the left). On the same circuit a global uniform
  foliation is represented as a set of thin staircase lines. The second circuit is obtained from the
  first one upon stretching wires in order to put the synchronous slices horizontally, so that the
  vertical axis represents time. The corresponding clock tic-tac is highlighted. The
  Lorentz space-contraction emerges as a consequence of the reduced density of events: the distance
  between mirrors changes from two to one events. Time-dilation is evident by counting the number
  of events during the complete tic-tac: eight on the left, sixteen on the right.
  \label{fig:synchronous}}
\end{figure}

In Ref. \cite{vax2009} I showed how Special Relativity emerges from the local causality of a
computational network. The basic idea is to make {\bf metric emerge from pure event-counting from the
purely topological structure of the computation}. The wires in the quantum circuit represent causal
connections, whence they can be stretched without affecting the information processing. Fig.
\ref{fig:synchronous} (from Ref. \cite{vax2009}) illustrates a factor-two Lorentz time-dilation and
space-contraction. 
The two reference-systems correspond to the usual horizontal foliation and
to the oblique staircase foliation drawn in the first circuit, respectively. Upon stretching the
wires in order to put the synchronous slices horizontally---so that the vertical axis represents
time---we obtain the second circuit, which is topologically identical to the first one.
Time-dilation and space-contraction emerge as a consequence of a resulting reduced density of
independent events, whereas time-dilation comes from a larger number of events during the same
tic-tac of an Einstein clock (light bouncing between two mirrors). With Alessandro Tosini we gave an
analytical derivation of the Lorentz transformations from a topologically uniform causal network
\cite{Lorentz}.  {\bf Topological uniformity plays the role of the Galileo principle}, \ie the invariance
of the physical law with the reference system---the physical law being the causal connection-rule of
the network, \ie a repeated tile of the pattern. Within a single frame the Galileo principle is just
uniformity of space and time: however, {\bf the causal network captures the conventionality of
homogeneity of space and time}---a long debated problem in Special Relativity (see \eg books
\cite{Brown,JammerSync})---while retaining the Galileo principle.
In a homogeneous causal network it is easy to see how causality is sufficient to guarantee a maximum
speed of information flow as one-event per step, corresponding to a line at $45^o$ in Fig.
\ref{fig:synchronous}. Simultaneity in relation to an observer is introduced as follows. {\bf The
observer is just an unbounded causal chain}. Thanks to the topological homogeneity, we can translate
the observer to any event in the network: simultaneity is defined by fixing an origin on the
observer chain, placing the origin of two identical translated observers over the two events,
sending light-signals from these events toward the other observer, and then comparing the number of
clock tic-tacs for each observer (for details see Ref. \cite{Lorentz}). Depending on the shape
of the observer chain, one may have situations in which there are no synchronous events.  However,
for periodic observers (each ziz-zag is a clock tic-tac) there always exist infinitely many
simultaneous events. The notion of simultaneity allows us to associate each observer with a
foliation of the causal network, and using the foliation on can define coordinates, by which the
Lorentz transformations are then derived.  It is worth re-emphasizing that the whole procedure
for defining space and time coordinates is made only by event-counting on the causal network.  An
essential ingredient in the transformation is a coarse-graining of events, corresponding to the
usual rescaling of Minkowski space to restore symmetry between the observers (this coarse-graining
is also related to the event-sparseness issue raised in Ref. \cite{Dowker} for Lorentz-transformed
regular lattices of points).

\section{The meaning of the inertial mass and of $\hbar$\label{s:mass}} \footnote{For this section  see also Ref. \cite{later}, which was written after the present manuscript was 
sent to press.}
As already said in the introduction, the information-theoretic meaning of inertial mass is simply
the slowing-down of the information flow due to repeated changes of direction, namely the Dirac's
Zitterbewegung. Thus, {\bf the inertial mass is the coupling between left and right propagating
  flows of information}. A coupling between left/and right propagating flows producing a periodic
oscillation of direction is described by the differential equation
\begin{equation}\label{zigzag}
\widehat\partial_t
\begin{bmatrix}\phi^+\\\phi^-\end{bmatrix}=
\begin{bmatrix}v\widehat\partial_x & -i\omega\\ -i\omega &
  -v\widehat\partial_x\end{bmatrix}\begin{bmatrix}\phi^+\\\phi^-\end{bmatrix}, 
\end{equation}
where $\phi^\pm$ denote the operators for the left/right propagating flows, $v$ is the speed of the
flow over the network, and $\widehat\partial_x$ and $\widehat\partial_t$ are finite-difference
derivatives. The coupling constant is denoted by the oscillation angular frequency $\omega$. Eq.
(\ref{zigzag}) has the same form of a Dirac equation (without spin), which means that the
quantum-information processing simulates a Dirac field.  This leads us to write the coupling
constant in terms of the Compton wavelength $\lambda$ as $\omega=c\lambda^{-1}$, and this allows us
to establish the following relation between the mass $m$ expressed in usual units (grams) and the
informational mass $\omega$
\begin{equation}\label{myGod}
m=\frac{\tau^2}{a^2}\hbar\omega=\frac{1}{c^2}\hbar\omega.
\end{equation}
In Eq. (\ref{myGod}) $a$ and $\tau$ denote the {\bf topon} and the {\bf chronon}, respectively---\ie
the conversion factors from pure event-counting to the usual units of space and time: in other
words, they are the minimum space and time distances between gates in the information-processing.
Eq. (\ref{myGod}) also provides a meaning for the Planck constant: {\bf $\hbar$ is the conversion
  factor in the equivalence between the informational notion in $\mathbf{sec^{-1}}$ and the
  customary notion in $\mathbf{Kg}$ of the inertial mass}.  It is worth noticing that Eq.
(\ref{myGod}) also corresponds to the usual Planck formula in terms of particle rest energy, which
are thus reinterpreted as the equivalence between the two notions of mass.

\section{The mass-dependent refraction index of vacuum\label{refrindx}} 
As mentioned in the introduction, the new QCFT has observational consequences. For example, {\bf the
request of unitariety and linearity in the field for the evolution lead to a mass-dependent vacuum
refraction index}. This is a general feature of QCFT, and does not depend on details of the quantum
circuit.  It is due to the fact that QCFT respects causality, namely no information can propagate at
a speed greater than $c$, whereas in a customary simulation by a quantum computer with gates
infinitesimally close to the identity, Lorentz covariance is recovered only in the continuum limit.
I will now briefly outline the derivation of the mass-dependent refraction index of Ref.
\ref{refrindx}. In the same reference you can also find a detailed evaluation for a specific quantum
circuit.

In the following we will use the notation $\phi(x)=[\phi^+(x),\phi^-(x)]^T$ for a generic spin-less
scalar field (either Boson or Fermion) in one space-dimension, $+$ ($-$) denoting the right- (left)
propagating components. The time evolution of the field is given by $\phi(x,t)=U_t\phi(x) U_t^\dag$,
where $U_t$ is the global unitary evolution of the field up to time $t$, and $\phi(x):=\phi(x,0)$.
One has $|\phi(x,t)\>=\phi(x,t)^\dag|0\>=U_t|\phi(x)\>=U_t\phi(x) U_t^\dag|0\>$ and the vacuum must
be left invariant, namely $U_t^\dag|0\>=|0\>$. 

\begin{figure}[b]
\includegraphics[width=.5\textwidth]{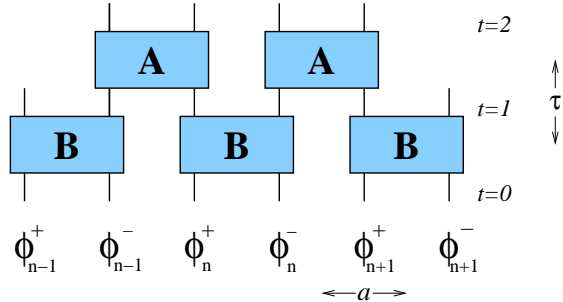}
\caption{General quantum circuit made only of bipartite gates for a massive field equation.}\label{f:KG}
\end{figure}

The discretized version of the field is given by $\phi_n:=a^{\frac{1}{2}}\phi(na)$, with
$n\in\mathbb{Z}$. In QCFT there is no Hamiltonian: all gates produce finite transformations, and the
unitary $U_t$ is the evolution due to all gates involved: in order to evaluate the evolution for few
chronons of a single system---\ie the field in a given location--- one needs to consider only a
finite set of gates. We can define an Hamiltonian in terms of a coarse-grained time-derivative of
the field as follows
\begin{equation}\label{Hdef}
\Hg{2k}z:=i\frac{z(k\tau)-z(-k\tau)}{2k\tau}=:i\widehat\partial_t z, 
\end{equation}
where $z$ denotes any field component at any given position, and the caret on the partial derivative
reminds us that it is discrete and coarse-grained. The field evolution is linear, whence we need to
consider gates that transforms the fields linearly. We consider just a single space-dimension, but
one can easily realize that the same reasoning apply to more dimensions. The coupling between left
and right-propagating field components corresponds to the following Dirac-like equation (modulo
unitarily equivalent representations)
\begin{equation}\label{dirac}
i\widehat\partial_t=c(i\zeta\sigma_z\widehat\partial_x+\lambda^{-1}\sigma_x), 
\end{equation}
where we consider a general velocity $v=\zeta c$ and we write the coupling frequency in the
convenient form $\omega=c\lambda^{-1}$. Using bipartite gates only, we need two rows of gates in the
forward direction to get a term $\phi_{n+1}^\varsigma$ ($\varsigma=\pm$) in the unitary evolution of
$\phi_n^\varsigma$, and similarly we need two rows of gates in the backward direction to get a term
$\phi_{n-1}^\varsigma$, namely we need $k=2$ in Eq.  (\ref{Hdef}). The general form of the circuit
is depicted in Fig. \ref{f:KG}.
The distance between the two adjacent positions $n$ and $n+1$ is $2a$ (a
right-propagating field covers it in two chronons), which means that the coarse-grained space
derivative to be considered is $\widehat\partial_x=\frac{1}{4a}(\delta_+ -\delta_-)$. This means
that in order to obtain Eq.  (\ref{dirac}) one must have the following overall backward and forward
unitary interactions
\begin{equation}\label{f-b}
U_f\phi_n^+U_f- U_b^\dag\phi_n^+ U_b=\zeta(\phi_{n+1}^+-\phi_{n-1}^+)-4i \frac{a}{\lambda}\phi_n^-.
\end{equation}
Therefore, since we consider gates involving only neighboring wires, the unitary evolutions must be
of the form
\begin{equation}
U_f\phi_n^+U_f^\dag=\eta\phi_n^++\zeta\phi_{n+1}^++\gamma\phi_n^-,\quad
U_b^\dag\phi_n^+U_b=\eta\phi_n^++\zeta\phi_{n-1}^++\gamma'\phi_n^-,
\end{equation}
with $\zeta>0$ and $\gamma-\gamma'=-4i \frac{a}{\lambda}$. The two evolutions may include other
terms that cancel each-other. But normalization of the row of the unitary matrix of the linear
evolution corresponds to
\begin{equation}
|\gamma|,|\gamma'|\leq\sqrt{1-\zeta^2}\;\Longrightarrow\;\frac{2a}{\lambda}\leq\sqrt{1-\zeta^2},
\end{equation}
which gives the lower bound for the vacuum refraction index
\begin{equation}
\zeta^{-1}\geq \sqrt{1-\left(\frac{2a}{\lambda}\right)^2}.
\end{equation}
At the end of the section I want to comment about the realization of field (anti)-commutations with
local Pauli-matrix algebras, \eg common qubits. The unitary transformation of each gate must
correspond to a unitary linear combinations of the field operators of the input wires. These can be
written in terms of local Pauli matrices using the Clifford algebraic construction
\begin{equation}
\phi_n^+=\sigma_{n+}^-\prod_{k=-\infty}^{n-1}\sigma_{k+}^z\sigma_{k-}^z,\quad
\phi_n^-=\sigma_{n-}^-\sigma_{n+}^z\prod_{k=-\infty}^{n-1}\sigma_{k+}^z\sigma_{k-}^z.
\end{equation}
Using the Fermionic Jordan-Schwinger realization of the unitary group $U(2)$ (of the unitary linear
combinations of the fields operators of the two wires at the gate), it is easy to see the unitary
transformation of gates are of the form
\begin{equation}
A=A(\phi_n^-,\phi_{n+1}^+)\equiv A(\vec\sigma_{n-},\vec\sigma_{(n+1)+}),\qquad
B=B(\phi_n^+,\phi_n^-)\equiv B(\vec\sigma_{n+},\vec\sigma_{n-}),
\end{equation}
namely each gate is a function only of the qubit algebra of its wires. The possibility of
generalizing the construction to more than one space-dimension, however, is still not obvious.

\section{Conclusions}
The new QCFT opens an extensive long-term research program. It represents the backbone of a complete
translation of Physics into information-theoretic terms. In addition to needing only QT without
quantization rules and giving Lorentz covariance for free, QCFT has the potential of solving many
problems that plague QFT \cite{vax2009}---being naturally a lattice theory---providing a new
unified framework for particle physics. 

The next immediate easy research steps will be to recover the commutation relations as emergent, to
retrieve the unitary representation of the Lorentz group, to classify the vacuum states, to
translate Lagrangian density into gates, to re-derive the Feynman's path integral. We need then to
generalize the information processing to more than 1 space dimension. Thereafter, less easy problems
need to be addressed, such as the issue of anticommuting fields (or parastatistics) and their
realization with local operators in more than one space dimensions. Then we want to exploit the
natural gauge invariance of the quantum circuit (local unitary transformations connecting the
Hilbert spaces of different wires in a slice) with the aim of deriving a first prototype of gauge
theory. Experimentally detectable effects will be the most interesting topic.

In the longer term we wants to explore how QCFT can help addressing the problem of Quantum Gravity.
There are two possible main paths: either we adopt a strong version of the equivalence
principle---in which the informational meaning of inertial mass coincides with the gravitational
one---or we adopt a causal set approach to gravity. In the first case gravity should emerge as a
quantum effect. In the second case we may need a non-causal variation of QT , \eg by considering an
information processing in a third-quantization fashion, where also the causal connections are
treated as ``states''. A possible route is to consider super-gates as the {\em switch-comb} of Ref.
\cite{switch}.

Whichever the research process will be, QCFT is a very promising testing ground for foundations of
QFT, an area of study that had never received the due attention, mostly due to the too technical
character of QFT.
 
\subsection*{Acknowledgments.} 
I thank Seth Lloyd, Masanao Ozawa, Paolo Perinotti, Alessandro Tosini, Lucien Hardy, Rafael Sorkin,
and Lee Smolin for very interesting discussions and suggestions. I'm grateful to researchers at the
National Labs of Frascati and at Fermilab of Chicago for useful feedback and very encouraging support.

\end{document}

%% file: myqcircuit.tex
\usepackage[matrix,frame,arrow]{xy}
\usepackage{amsmath}

\newcommand{\qw}[1][-1]{\ar @{-} [0,#1]}
\newcommand{\gate}[1]{*{\xy *+<.6em>{#1};p\save+LU;+RU **\dir{-}\restore\save+RU;+RD **\dir{-}\restore\save+RD;+LD **\dir{-}\restore\POS+LD;+LU **\dir{-}\endxy} \qw}
\newcommand{\multimeasureD}[2]{*+<1em,.9em>{\hphantom{#2}}\save[0,0].[#1,0];p\save !C *{#2},p+LU+<0em,0em>;+RU+<-.8em,0em> **\dir{-}\restore\save +LD;+LU **\dir{-}\restore\save +LD;+RD-<.8em,0em> **\dir{-} \restore\save +RD+<0em,.8em>;+RU-<0em,.8em> **\dir{-} \restore \POS !UR*!UR{\cir<.9em>{r_d}};!DR*!DR{\cir<.9em>{d_l}}\restore \qw}
\newcommand{\multigate}[2]{*+<1em,.9em>{\hphantom{#2}} \qw \POS[0,0].[#1,0];p !C *{#2},p \save+LU;+RU **\dir{-}\restore\save+RU;+RD **\dir{-}\restore\save+RD;+LD **\dir{-}\restore\save+LD;+LU **\dir{-}\restore}
\newcommand{\ghost}[1]{*+<1em,.9em>{\hphantom{#1}} \qw}
\newcommand{\ustick}[1]{*!D!<0em,-.5em>=<0em>{#1}}
\newcommand{\Qcircuit}[1][0em]{\xymatrix @*[o] @*=<#1>}
\newcommand{\pureghost}[1]{*+<1em,.9em>{\hphantom{#1}}}
\newcommand{\multiprepareC}[2]{*+<1em,.9em>{\hphantom{#2}}\save[0,0].[#1,0];p\save !C
  *{#2},p+RU+<0em,0em>;+LU+<+.8em,0em> **\dir{-}\restore\save +RD;+RU **\dir{-}\restore\save
  +RD;+LD+<.8em,0em> **\dir{-} \restore\save +LD+<0em,.8em>;+LU-<0em,.8em> **\dir{-} \restore \POS
  !UL*!UL{\cir<.9em>{u_r}};!DL*!DL{\cir<.9em>{l_u}}\restore}
\newcommand{\prepareC}[1]{*{\xy*+=+<.5em>{\vphantom{#1\rule{0em}{.1em}}}*\cir{l^r};p\save*!L{#1} \restore\save+UC;+UC+<.5em,0em>*!L{\hphantom{#1}}+R **\dir{-} \restore\save+DC;+DC+<.5em,0em>*!L{\hphantom{#1}}+R **\dir{-} \restore\POS+UC+<.5em,0em>*!L{\hphantom{#1}}+R;+DC+<.5em,0em>*!L{\hphantom{#1}}+R **\dir{-} \endxy}}

%% file: Dirac_vaxjo2010.bbl
\begin{thebibliography}{0}
\bibitem{Hardy} L. Hardy, \emph{Quantum theory from five reasonable axioms}, quant-ph/0101012v4 (2001).
\bibitem{Fuchs} C. A. Fuchs, {\em Quantum Mechanics as Quantum Information (and only a
    little more)}, quant-ph/0205039 (2002).
\bibitem{my2005} G. M. D'Ariano, \emph{On the missing axiom of quantum mechanics},
  Quantum Theory, Reconsideration of Foundations - 3, V\H{a}xj\H{o}, Sweden, 6-11 June 2005
  (Melville, New York) (G.~Adenier, A.~Y. Khrennikov, and T.~M. Nieuwenhuizen, eds.), (AIP,
  Melville, New York 2006) pp.~114.
\bibitem{myCUP2009} G. M. D'Ariano, in \emph{Philosophy of quantum information and entanglement},
  A.~Bokulich and G.~Jaeger eds., (Cambridge University Press, Cambridge UK, 2010). Also arXive
  0807.4383.
\bibitem{CDP2010} G. Chiribella, G. M. D'Ariano, P. Perinotti, {\em Probabilistic Theories
    with Purification}, Phys. Rev. A  {\bf 81} 062348 (2010).
\bibitem{ournature} G. Chiribella, G. M. D'Ariano, P. Perinotti, {\em Quantum Theory is a Theory of Information}
  (work in progress: see arXiv 2010-11)
\bibitem{support} G. Chiribella, G. M. D'Ariano, P. Perinotti, {\em Informational derivation of Quantum
    Theory} arXiv 2011.6451 (2010).
\bibitem{Bombelli-Sorkin_(1987)} L.~Bombelli, J.~H.~Lee, D.~Meyer, and R.~Sorkin, {\em Space-Time as
    a Causal Set}, Phys. Rev. Lett. {\bf 59}, 521 (1987). 
\bibitem{Blute} R. Blute, I. Ivanov, and P. Panangaden, {\em Discrete quantum causal dynamics},
Int. J. Theor. Phys. {\bf 42} 2025 (2003)
\bibitem{Hardy:2009} L. Hardy, {\em Foliable Operational Structures for General Probabilistic Theories},
arXiv: 0912.4740 (2009)
\bibitem{vax2009} G. M. D'Ariano, {\em On the "principle of the quantumness", the quantumness of
    Relativity, and the computational grand-unification}, in CP1232 {\em Quantum Theory:
    Reconsideration of Foundations, 5}, edited by A.~Y. Khrennikov (AIP, Melville, New York, 2010),
  pag 3. (Also arXiv:1001.1088)
\bibitem{Knuth} K.~H.~Knuth, N.~Bahreyni, arXiv: 1005.4172 (2010).
\bibitem{Lorentz} G. M. D'Ariano and A. Tosini, {\em Space-time and special relativity from causal
    networks}, arXiv: 1008.4805 (2010)
\bibitem{WolframBook} S. Wolfram. {\em A New Kind of Science}, Wolfram Media (Champaign, 2002).
\bibitem{lamport} L. Lamport, {\em Time, clocks, and the ordering of events in a distributed
    system}, Comm. ACM, {\bf 21} 558 (1978)
\bibitem{stein} I. Stein, {\em The Concept of Object As the Foundation of Physics}, San Francisco
  State University Series in Philosophy, vol. 6 (Peter Lang Publishing, NY 1996) 
\bibitem{Bialynicki} I. Bialynicki-Birula, {\em Weyl, Dirac, and Maxwell equations on a lattice as
    unitary cellular automata},  Phys. Rev. D {\bf 49} 6920 (1994)
\bibitem{Boghosian} B. M. Boghosian and W. Taylor, {\em Simulating quantum mechanics on a quantum
    computer}, Physica D: Nonlinear Phenomena {\bf 120} 30 (1998)
\bibitem{Meyer} D.A. Meyer, {\em From quantum cellular automata to quantum lattice gases},
J. Stat. Phys. {\bf 85} 551 (1996)
\bibitem{toytheories} G. M. D'Ariano and A. Tosini, {\em Testing axioms for Quantum Theory 
on Probabilistic toy-theories}, Quant. Inf. Proc. {\bf 9} 95-141 (2010) (Special Issue on
Foundations of Quantum Information) (also arXiv:0911.5409)
\bibitem{Brown} Harvey R. Brown, {\em Physical Relativity: Space-Time Structure from a Dynamical
    Perspective} (Clarendon Press, Oxford, 2005)
\bibitem{JammerSync} Max Jammer, {\em Concepts of Simultaneity: From Antiquity to Einstein and
    Beyond} (John Hopkins University, Baltimore Press 2006)
\bibitem{later} G. M. D'Ariano, {\em The Quantum Field as a Quantum Computer}, arXiv (2010)
\bibitem{Dowker} F.~Dowker, J.~Henson, and R.~D.~Sorkin, Mod. Phys. Lett. A {\bf 19} 1829 (2004).
\bibitem{switch} G. Chiribella, G. M. D'Ariano, P. Perinotti, and B. Valiron, {\em Beyond Quantum
  Computers}, arXiv:0912.0195 
\end{thebibliography}
